\documentclass[preprint]{revtex4-1}

\usepackage[T1]{fontenc}
\usepackage[utf8]{inputenc}
\usepackage[english]{babel}
\usepackage{lmodern}
\usepackage{amsmath}
\usepackage{amssymb}
\usepackage{latexsym}
\usepackage{graphicx}

\providecommand{\vr}{\mathbf{r}}
\renewcommand{\vr}{\mathbf{r}}
\newcommand{\vG}{\mathbf{G}}
\newcommand{\vk}{\mathbf{k}}
\newcommand{\rd}{\mathrm{d}}
\newcommand{\ii}{\mathrm{i}}
\newcommand{\ee}{\mathrm{e}}
\newcommand{\rA}{\mathrm{A}}
\newcommand{\rB}{\mathrm{B}}

\begin{document}

\date{\today}
\title{Inter-ribbon tunneling in graphene: an atomistic Bardeen approach}
\author{Maarten L. \surname{Van de Put}}
\email[Electronic mail: ]{maarten.vandeput@uantwerpen.be}
\affiliation{Department~of~Physics, Universiteit~Antwerpen, Belgium}
\affiliation{imec, Kapeldreef~75, 3001~Leuven, Belgium}
\author{William G. \surname{Vandenberghe}}
\affiliation{Department of Material Science, University of Texas at Dallas, Texas, USA}
\author{Bart \surname{Sor\'ee}}
\affiliation{Department~of~Physics, Universiteit~Antwerpen, Belgium}
\affiliation{imec, Kapeldreef~75, 3001~Leuven, Belgium}
\author{Wim \surname{Magnus}}
\affiliation{Department~of~Physics, Universiteit~Antwerpen, Belgium}
\affiliation{imec, Kapeldreef~75, 3001~Leuven, Belgium}
\author{Massimo V. \surname{Fischetti}}
\affiliation{Department of Material Science, University of Texas at Dallas, Texas, USA}

\begin{abstract}
A weakly coupled system of two crossed graphene nanoribbons exhibits direct tunneling due to the overlap of the wavefunctions of both ribbons.
We apply the Bardeen transfer Hamiltonian formalism, using atomistic band structure calculations to account for the effect of the atomic structure on the tunneling process.
The strong quantum-size confinement of the nanoribbons is mirrored by the one-dimensional character of the electronic structure, resulting in properties that differ significantly from the case of inter-layer tunneling, where tunneling occurs between bulk two-dimensional graphene sheets.
The current-voltage characteristics of the inter-ribbon tunneling structures exhibit resonance, as well as stepwise increases in current. 
Both features are caused by the energetic alignment of one-dimensional peaks in the density-of-states of the ribbons.
Resonant tunneling occurs if the sign of the curvature of the coupled energy bands is equal, whereas a step-like increase of the current occurs if the signs are opposite.
Changing the doping modulates the onset-voltage of the effects as well as their magnitude.
Doping through electrostatic gating makes these structures promising for application towards steep slope switching devices. 
Using the atomistic empirical pseudopotentials based Bardeen transfer Hamiltonian method, inter-ribbon tunneling can be studied for the whole range of two-dimensional materials, such as transition metal dichalcogenides.
The effects of resonance and of step-like increases of the current we observe in graphene ribbons, are also expected in ribbons made from these alternative two-dimensional materials, because these effects are manifestations of the one-dimensional character of the density-of-states.
\end{abstract}

\maketitle

\section{Introduction}

Van der Waals heterostructures of two-dimensional (2D) materials have seen increasing research interest.\cite{Geim:2013hf}
Their unique 2D electronic properties and atomically perfect interfaces yield interesting new physical effects.
For example, 2D heterostructures can be used for transistor applications by relying on interlayer tunneling\cite{Feenstra:2012ge}, Bose-Einstein condensation of spatially indirect excitons\cite{banerjee2009bilayer}, or topological insulating properties\cite{vandenberghe2014calculation}.
Specific materials currently of interest are transition-metal dichalgenides\cite{wang2012electronics}, phosphorene\cite{liu2014phosphorene} and stanene\cite{xu2013large}, but graphene, the `first' 2D material that has risen to widespread interest, and its heterostructures have been studied experimentally and theoretically the most.

The rotational alignment of two graphene layers shows significant impact on the electronic structure near the Dirac cones. Most interestingly, even a slight rotation of the layers causes a twofold increase of the degeneracy of the Dirac cone energy as the Dirac cones of the two constituent layers separate due to a rotation of their Brillouin zones.\cite{Shallcross:2009fs}
Due to this change in electronic structure, the conductivity of the graphene stack is highly dependent on the relative orientation of the individual graphene sheets\cite{Bistritzer:2010kl}.
Furthermore, in the case of high misalignment, it is shown both theoretically and experimentally that electron transport between the layers is dominated by phonons, which provide the momentum necessary to transition between the shifted Dirac cones of the individual sheets.\cite{Perebeinos:2012kb,Habib:2012kdb,Kim:2013fe}

The introduction of an insulating layer, such as hexagonal boron-nitride (hBN), yields a graphene-insulator-graphene (G-I-G) stack with similar electronic properties.
Theoretical study of the current-voltage characteristics of these G-I-G stacks with misaligned Dirac cones predicts negative differential resistance (NDR)\cite{Feenstra:2012ge,delaBarrera:2014cy,Lebedeva:2015hp,Zhao:2015ci} as well as flat regions in the current\cite{Roy:2014cx}.
Recently, resonant tunneling has been measured between two rotationally misaligned graphene sheets separated by an hBN layer\cite{Mishchenko:2014hu}.
Further experiments also confirm NDR in aligned G-I-G structures with different chemical doping of the layers\cite{Fallahazad:2015fy}.
In both cases the observed NDR is caused by shifting the Dirac cones in energy by applying a bias voltage. 
When properly aligned, the electrons directly transition between the states of both layers with momentum and energy conserved without the aid of phonons. This allows for a high direct tunneling current at resonance.

The inter-ribbon tunneling current has almost universally been described using the Bardeen transfer Hamiltonian method\cite{Bardeen:1961hd}.
This method was first introduced by Bardeen to describe tunneling from a many-particle point of view between two metals separated by an oxide and yields a good first-order approximation of the tunneling current between two weakly-coupled systems and is now commonly used to describe the tip-to-sample current in scanning tunneling microscopy (STM)\cite{Gottlieb:2006fy,Noguera:1989gu,Tersoff:1983id}.
While the Bardeen transfer Hamiltonian requires weak coupling between subsystems this does not automatically restrict the tunneling probabilities to be low, as is evidenced in the case of resonant tunneling\cite{Payne:1986gw}.
These features, combined with ease of use make it the preferred method for inter-layer tunneling.

\begin{figure}[h!]
  \includegraphics[width=.46\textwidth]{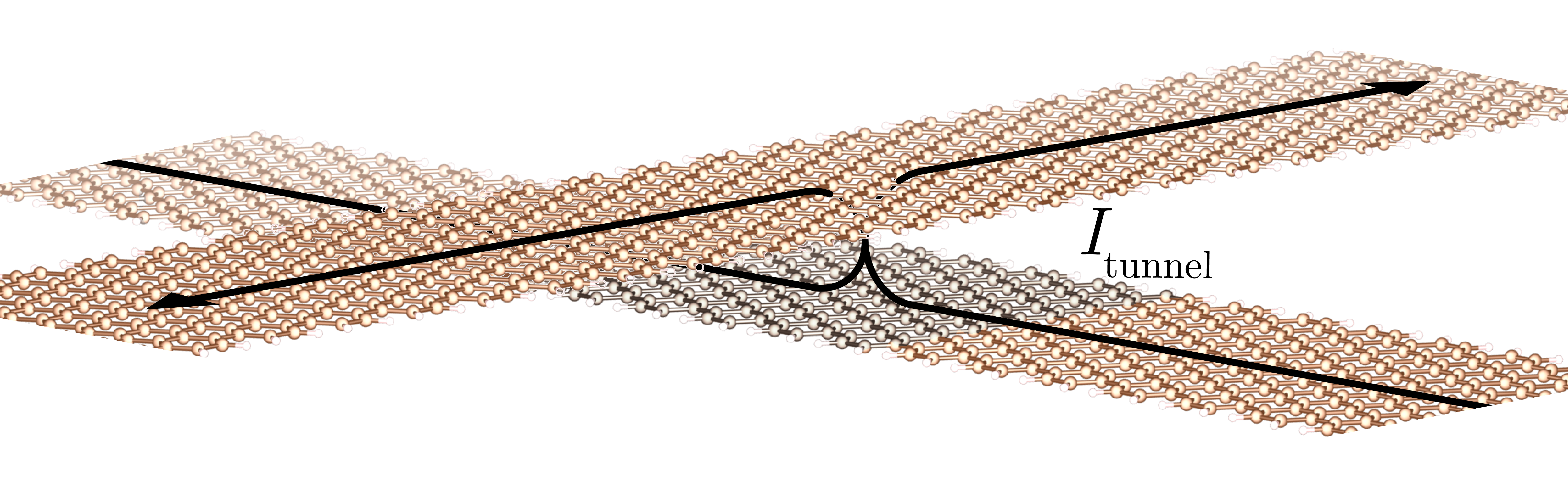}
  \caption{Illustration of an inter-ribbon tunneling structure of two crossed graphene nanoribbons where the inter-ribbon separation has been exaggerated for clarity.. Current flows from one ribbon by tunneling in the overlap region, as indicated. The depicted ribbons have an armchair edge and are only $2$ nm wide.}
  \label{f:atomistic-current}
\end{figure}
Contrary to inter-layer tunneling, which relies on ``bulk'' 2D properties, conservation of 2D momentum is not required in nanoscaled structures such as nanoribbon tunneling structures (illustrated in Fig.~\ref{f:atomistic-current}). 
Studying nanoscale structures is important from an application point of view and crossed graphene ribbons and carbon nanotubes have been fabricated and measured.\cite{Jiao:2010jn}
Moreover, the properties of graphene ribbons are strongly dependent on edge type and width, yielding a large parameter space for inter-ribbon tunneling structures with potentially interesting effects and technological applications.
Due to their intricate electronic structure, detailed, atomistic methods are required to theoretically model the tunneling current between graphene nanoribbons.

The electronic structure of nanoscaled tunneling devices has been described using k$\cdot$p, tight-binding (TB) and density functional theory (DFT)\cite{Gonzalez:2010do,Gonzalez:2011gx,Habib:2013dj,Habib:2012kda}.
These studies show that resonant tunneling in the form of NDR is also present in nanoscaled structures.
In contrast to the Bardeen transfer Hamiltonian approach, these studies calculated the tunneling current using the non-equilibrium Green's function (NEGF) approach.
However, Habib et al. showed that the NEGF approach they used matches well with a transfer Hamiltonian description\cite{Habib:2013dj}.

In this paper, we compute the current between two graphene ribbons in an atomistic way. 
We use empirical pseudopotentials to compute the bandstructure atomistically and the Bardeen transfer Hamiltonian formalism to calculate the direct tunneling current. 
Empirical pseudopotentials offer atomistic accuracy at a low computational cost without the need to introduce \emph{ad hoc} parameters.
We use the transfer Hamiltonian formalism as opposed to a more complex model such as NEGF to describe the relevant physics because it facilitates interpretation while reducing the computational burden.
Additionally, the NEGF formalism requires the introduction of contacts, whereas in most cases we are interested in the intrinsic tunneling process occurring between nanoribbons, and not in the specific effects related to the contacts.
Using this formalism, we calculate the tunneling current between ribbons of several sizes and different edge types, and trace back the physical origins of the observed behavior.
We find that the one-dimensional nature of the bandstructure defines the tunneling current through the peaks in the density of states, leading to both resonance and sharp current changes.
We only model graphene nanoribbons, but we expect our conclusions can be generalized to ribbons made from other 2D materials, since the basic physics is similar and our computational method is no less applicable.

In section~\ref{modeling}, we introduce the atomistic Bardeen transfer Hamiltonian method used to model the inter-ribbon tunneling process.
In section~\ref{electronic_structure}, we consider the obtained electronic structure of the different ribbons, with the intent of highlighting the key features needed to understand inter-ribbon tunneling.
In section~\ref{tunneling_current}, we calculate the inter-ribbon tunneling current for different combinations of ribbons, and explain the origin of the observed effects using the electronic structure and Bardeen transfer Hamiltonian method.
Finally, we conclude in section~\ref{conclusions}.

\section{Modeling}
\label{modeling}

We model the behavior of the crossed two-ribbon system in two steps.
First, we treat the two ribbons as separate systems, based on the inherently weak coupling between the two ribbons,
and we determine the electronic structure for each ribbon using empirical pseudopotentials.
In a second step, we determine the direct inter-ribbon tunneling from the electronic structure using the Bardeen transfer Hamiltonian method.

\subsection{Empirical pseudopotentials}
\label{modeling:ep}

The electronic structure of the ribbons is determined atomistically using local, empirical pseudopotentials.
This approximation has been shown to be well suited for the calculation of the electronic structure of graphene nanoribbons.\cite{Fischetti:2011he,Fischetti:2013wi,Fischetti:2013zi}
Excellent empirical pseudopotentials parameters are available to characterize most carbon allotropes.\cite{Kurokawa:2000cw}

Empirical pseudopotentials are intended to correctly represent the ionic potential outside of the core region, thus providing high accuracy for wavefunctions outside the core region and with energies close to the Fermi level.
The accuracy of the wavefunctions in this region is important to calculate the inter-ribbon tunneling current correctly using the Bardeen transfer Hamiltonian method.

We solve the empirical pseudopotential Hamiltonian by discretizing on a plane-wave basis set with reciprocal lattice vectors ${\vG}$. The Hamiltonian is then given by
\begin{equation}
  H_{\vG\vG'}(\vk) = \frac{\hbar^2|\vG+\vk|^2}{2m} \delta_{\vG\vG'} + \tilde{V}_{\rm ep}(|\vG - \vG'|)\,,
\end{equation}
where $\vk$ is the wavevector and $\tilde{V}_{\rm ep}(q)$ the reciprocal pseudopotential of all atoms in a supercell.
The  wavefunctions $\psi_{n,\vG}(\vk)$ are the eigenfunctions of the Hamiltonian solving the Schr\"odinger equation
\begin{equation}
  H_{\vG\vG'}(\vk) \psi_{n,\vG}(\vk) = E_n(\vk) \psi_{n,\vG}(\vk)\,.
\end{equation}
Directly solving this equation requires the diagonalization of a full-rank matrix due to the plane-wave expansion of the empirical pseudopotential.

To alleviate the computational burden associated the direct diagonalizing of a big full rank matrix using standard routines, we  evaluate the potential energy in real space performing Fourier transforms $\mathcal{F}$, as the potential energy term is responsible for the matrix being dense in the plane wave expansion.
The Schr\"odinger equation then reads
\begin{multline}
  \frac{\hbar^2|\vG+\vk|^2}{2m} \psi_\vG(\vk)
  + \mathcal{F}^{-1}\!\left\{ V_{\rm ep}(\vr) \mathcal{F}\left[\psi_{\vG}(\vk)\right]\!(\vr) \right\}_{\vG} \\
  = E(\vk)\, \psi_\vG(\vk)\,.
  \label{e:ep_schroedinger}
\end{multline}
The evaluation of the potential energy is then also a diagonal operation, at the cost of two Fourier transforms.
However, by using a highly optimized Fast Fourier Transform (FFT) algorithm\cite{FFTW05} we have reduced the numerical complexity from $\mathcal{O}(n^2)$ for the direct evaluation of the Hamiltonian to $\mathcal{O}(n\log n)$ for the real space method, where $n$ is the size of the numerical problem, i.e. the total number of plane wave components in each direction. For given computational resources (CPU, working memory and time), the FFT based method will be able to handle much bigger structures which have a large number of required plane wave components than direct matrix implementation would allow.

Due to the lack of a matrix representation, solving the eigenvalue system using the real-space, fft-based method requires a matrix-free method, which eliminates direct numerical solvers.
We have implemented the highly efficient ``residual minimization method by direct inversion of the iterative subspace'' (RMM-DIIS) eigenvalue algorithm\cite{Wood:1985de,Kresse:1996kl} to solve Eq.~(\ref{e:ep_schroedinger}).
The RMM-DIIS algorithm is an iterative method that directly optimizes for the eigenfunctions (wavefunctions) by minimizing the residual using an accelerated steepest descent procedure, thereby allowing for the semi-independent evaluation of the different eigenfunctions i.e. bands.
Carrying out calculations for different $\vk$-points and different band indices $n$ in parallel, the solver easily scales to larger ribbon sizes while maintaining short computational times.
Our empirical pseudopotential solver is input/output-compatible with the Vienna ab-initio package (VASP) as it accepts POSCAR and KPOINTS files and outputs the bandstructure in WAVECAR format for convenience.\cite{Kresse:1994us,Kresse:1993ty}

In the case of graphene ribbons, solving Eq.~(\ref{e:ep_schroedinger}) yields the dispersion relation $E_n(k_z)$ for band index $n$ and $k_z$, the $z$ component of the wavevector, as well as the wavefunctions of each state $\psi_{n,\vG}(k_z)$.

\subsection{Bardeen transfer Hamiltonian}

Using the Bardeen approximation, the total Hamiltonian of the inter-ribbon system comprised of ribbons $\rA$ and $\rB$ is expressed in terms of the Hamiltonians of the separate ribbons $H_\rA$ and $H_\rB$ and a transfer Hamiltonian $H_\mathrm{T}$,
\begin{equation}
H = H_\rA + H_\rB + H_\mathrm{T}\,,
\end{equation}
where, by inspection, the transfer Hamiltonian is the negative of the kinetic energy $H_\mathrm{T}=-T$.

The Bardeen transfer Hamiltonian approach follows from first order time-dependent perturbation theory, the transfer Hamiltonian being a perturbation that neglects two types of matrix elements; (1) the overlaps between wavefunctions of disparate ribbons, and (2) the matrix elements of the wavefunction of one ribbon with the crystal potential of the other ribbon, together consituting the weak-coupling assumption.
The only remaining coupling terms are the transfer Hamiltonian matrix elements, \emph{i.e.}, the kinetic energy matrix elements between wavefunctions of different ribbons,
\begin{equation}
  T_{nn'}(k_z, k_z')
  = \int \rd^3r\, \phi^{\rA*}_{n,k_z}(\vr)\, \mathrm{T}\, \phi^\rB_{n',k_z'}(\vr)\,,
  \label{e:matrix}
\end{equation}
where the real space wavefunctions are expanded as
\begin{equation}
  \phi_{n,k_z}(\vr) = \sum_\vG \psi_{n,\vG}(k_z) \ee^{\ii(\vG+k_z \hat{z})\cdot\vr}\,,
\end{equation}
and the kinetic energy operator is applied in the momentum basis $\mathrm{T} = \hbar^2|\vG|^2/2m$.
In the standard Bardeen approach, the volumetric kinetic matrix elements are further transformed into ``flux'' matrix elements of the probability current operator normal to a plane between the two sub-systems, by means of the divergence theorem. 
We have elected to work with the kinetic matrix elements directly to keep the numerical evaluation simple and less susceptible to numerical errors.

Fermi's golden rule yields the first-order approximation to the direct inter-ribbon tunneling current,
\begin{multline}
  I = \frac{2\pi\ee}{\hbar} \,
      \sum_{nn'} \int\rd k \int\rd k'\,
       \big|T_{nn'}(k, k')\big|^2\\
       \Big\{f\big[E_{n}(k)-\mu_\rA\big] - f\big[E_{n'}(k')-\mu_\rB\big]\Big\} \\
       \delta\big[ E_{n}(k)-E_{n'}(k') - \ee V_\mathrm{ds}\big]\,.
  \label{e:current}
\end{multline}
with $V_\mathrm{ds}$ the applied bias and the separate Fermi levels $\mu_\rA$ and $\mu_\rB$.

To compute the matrix elements, we import the wavefunctions obtained from our empirical pseudopotential solver and compute the matrix element of the kinetic energy operator $\mathrm{T}$.
For efficiency, we evaluate the kinetic energy in reciprocal space.
In transforming to real space, we account for the positions of the individual ribbons by translation using a phase factor $\ee^{\ii \vG\cdot\mathbf{\Delta}}$, where $\Delta$ is the position offset, and rotating in real space.

The numerical evaluation of the current involves a double integration of energy-conserving delta functions.
Given a fine $k$ grid, the Dirac-delta function is well approximated by a Gaussian $\exp[-E^2/(2\Delta E^2)]$.
However, because the calculation of the matrix elements at many $k$-points is computationally intensive, and the kinetic matrix elements $T_{nn'}(k_z, k_z')$  do not vary appreciably for small changes of their wavevector arguments, we calculate the matrix elements for a limited set of $k$-points and perform a bilinear interpolation between them.
For the bandstructure, linear integration is not sufficient, as it does not properly capture the high density-of-states (DoS) at band extrema. 
Fortunately, we can readily calculate the first derivative 
$\rd E(k)/\rd k_z = \hbar^2/m \sum_\vG ((0,0,k_z)+\vG)\, \psi_\vG^*\psi_\vG$
by invoking the Hellmann–Feynman theorem.\cite{Feynman:1939bg,Hellmann:1937wb}
With this information, we approximate $E(k)$ with a piecewise quartic function, properly capturing the band extrema at much lower computational cost.

In our simulations, we have found that to compute the electronic structure, $20$ $k$-points are sufficient.
Taking the momentum matrix element between each state, yields a total of $400$ matrix elements for every two bands.
The Bardeen integral is then calculated using the interpolations on an ever finer grid until the result reaches convergence.

Inspection of Eq.~(\ref{e:current}) reveals three factors contributing to the current with different physical origins. 
The first factor arises from the kinetic energy matrix elements that determine the strength of coupling and is the only factor that is directly influenced by the geometry of the ribbons. 
The second contribution comes from the distribution functions, which are determined by the doping level and are susceptible to gating effects.
Lastly, we have the density of states present in Eq.~(\ref{e:current}) as a sum over the energy conserving delta functions.
The density of states is fully determined by the type and width of the individual ribbons.

\section{Electronic structure}
\label{electronic_structure}

Using the model described in section~\ref{modeling:ep}, we calculate the electronic properties of a range of graphene ribbons.
Our aim in this section is twofold: we verify our model of the electronic structure by reproducing expected results and additionally, we highlight the features in the electronic structure that are relevant in the inter-ribbon tunneling process. 
We position the atoms of the first and the second ribbon each in a periodic supercell, where the distance between carbon atoms is $1.42\AA$. The structure is hydrogen terminated with a bond length of $1.09\AA$. The inter-atomic distances have been taken from relaxed DFT calculations in small graphene ribbons using VASP.\cite{Kresse:1994us,Kresse:1993ty}
The ribbons are assumed to be periodic in their continuous $z$ direction.
We add vacuum spacing of  $2$ nm in the perpendicular $x-y$ directions to avoid erroneous inter-ribbon coupling and to correctly account for the decay of the wavefunctions into vacuum.
The empirical pseudopotential parameters for carbon and hydrogen are taken from Kurokawa et al.\cite{Kurokawa:2000cw}

We consider two ribbon types, the first having a zigzag edge, while the second has an armchair edge.
In the following, we show the results for both types, highlighting the properties that are key to understanding the inter-ribbon tunneling process. 

\subsection{Armchair edge}

\begin{figure}[h!]
  \includegraphics{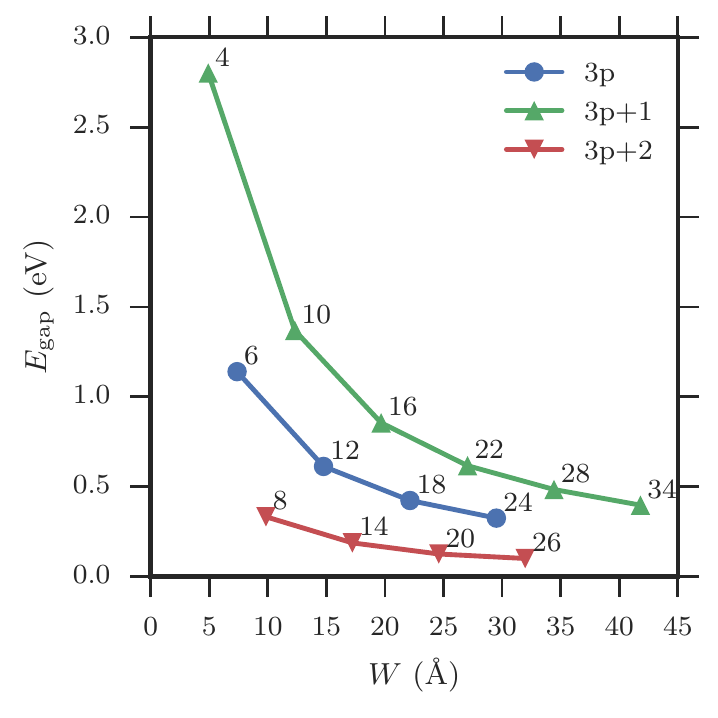}
  \caption{Dependence of armchair nanoribbon bandgap $E_{\rm gap}$ on the ribbon width $W$. Three distinct categories emerge based on the number of carbon lines (annotations) $3p$, $3p+1$ or $3p+2$.}
  \label{f:armchair_bandgaps}
\end{figure}

We begin with the armchair-edge ribbons, where different Clar resonance structures are possible for different ribbon widths leading to a categorization in three Claromatic classes.\cite{Clar:1972ut,Balaban:2009gr,Fischetti:2013zi}
The classes are determined by the number of parallel carbon-lines; either $3p$, $3p+1$ or $3p+2$, where $p$ a positive integer.
While the properties of the ribbons within one class are similar, they may substantially vary between ribbons of different claromatic classes, even for small changes in ribbon width.
This classification of the armchairs is visible in the trend of the bandgaps with increasing ribbon width, as illustrated in Fig.~\ref{f:armchair_bandgaps}.
The armchair ribbons all have a bandgap and are thus semiconducting.

\begin{figure}[h!]
  \includegraphics{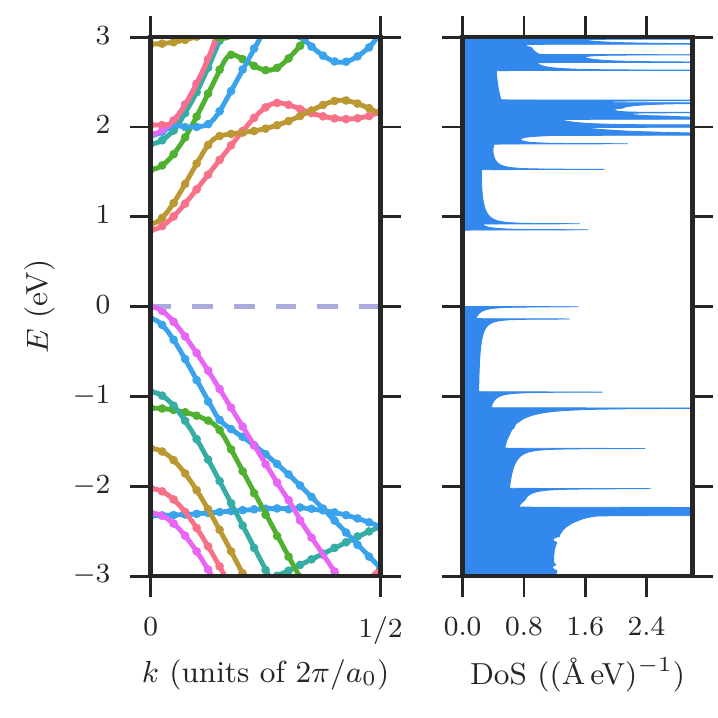}
  \caption{The bandstructure and density-of-states of a 16 C-atom-wide armchair ribbon. Energies are referenced to the Fermi level at $0$ eV.}
  \label{f:armchair_16_bsdos}
\end{figure}

\begin{figure}[h!]
  \includegraphics{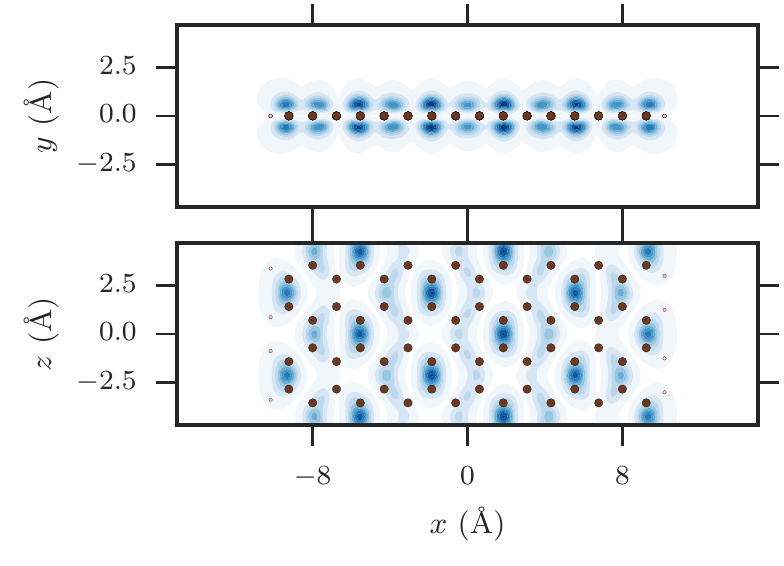}
  \caption{Squared amplitude of the first valence-band state of a 16 C-atom-wide armchair ribbon in the supercell, averaged in the $z$ direction (top) and the $y$ direction (bottom).}
  \label{f:armchair_16_valence}
\end{figure}

\begin{figure}[h!]
  \includegraphics{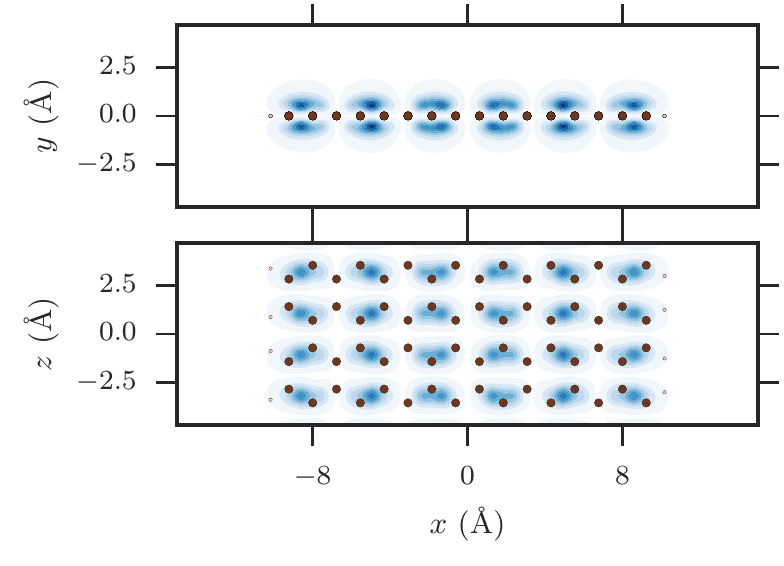}
  \caption{Squared amplitude of the first conduction band state of a 16 C-atom-wide armchair ribbon in the supercell, averaged in the $z$ direction (top) and the $y$ direction (bottom).}
  \label{f:armchair_16_conduction}
\end{figure}

As a typical example, we discuss the electronic structure of a 16 carbon lines wide ($3p+1$) armchair-edge ribbon.
The ribbon is approximately $2$ nm wide.
In Fig.~\ref{f:armchair_16_bsdos} we plot its bandstructure near the Fermi level (arbitrarily set at $0$ eV), showing a bandgap of $0.9$ eV.
The corresponding density of states exhibits the typical $1/\sqrt{E}$ peaks of a one-dimensional structure.
The first conduction and valence-band states at $k=0$ are important in the tunneling process.
In figures~\ref{f:armchair_16_valence} and \ref{f:armchair_16_conduction} we show the probability density profile of these states.
The states are distributed throughout the entire ribbon, which clearly shows they are bulk states.

\subsection{Zigzag edge}

In the local empirical pseudopotential approach, all zigzag edge-states are metallic in contrast to the armchair-edge ribbons.
But more advanced band structure calculation methods, accounting for the anti-ferromagnetic (AF) coupling between opposite edges\cite{Son:2006ky,Jung:2009ds,Yang:2007bx}, reveal the opening of a small energy bandgap.
For our purpose, the use of local empirical pseudopotentials is satisfactory since the bandgap induced by the AF coupling rapidly becomes small for increasing ribbon widths.

\begin{figure}[h!]
  \includegraphics{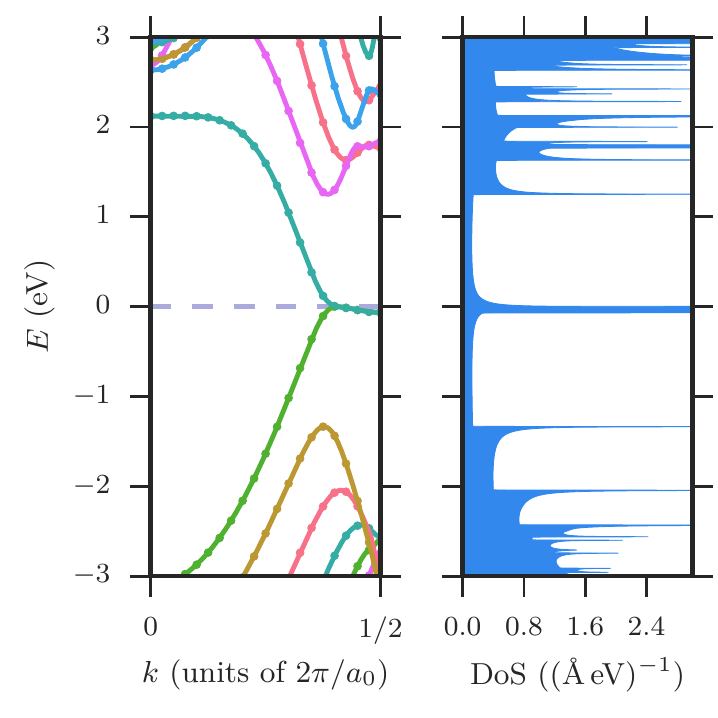}
  \caption{The bandstructure and density-of-states of a 8 C-atom-wide zigzag ribbon. Energies are referenced to the Fermi level.}
  \label{f:zigzag_08_bsdos}
\end{figure}

\begin{figure}[h!]
  \includegraphics{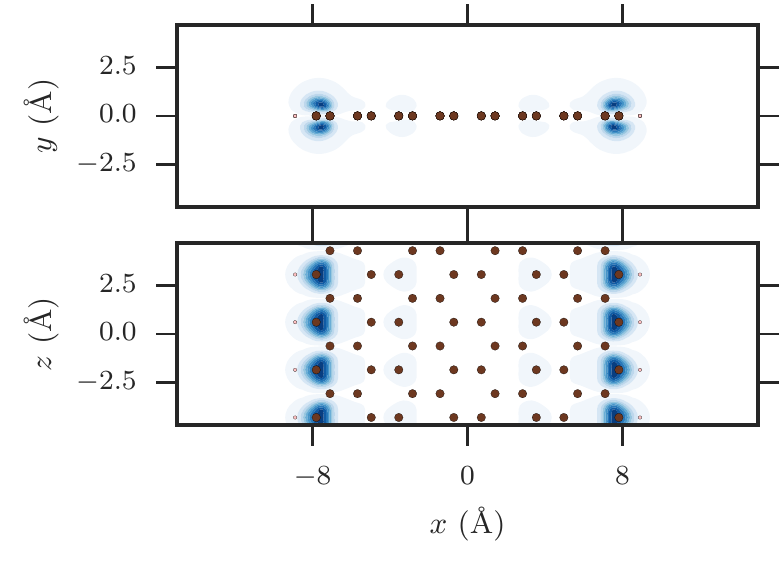}
  \caption{Squared amplitude of the states of a 8 C wide zigzag ribbon at $k=G_0/2$. Only one state is shown as both states have the same squared amplitude with opposite phase symmetry.}
  \label{f:zigzag_08_valence}
\end{figure}

Once again we present one ribbon as a typical example of zigzag-edge ribbons in Fig.~\ref{f:zigzag_08_bsdos}; specifically, we show an 8 carbon-atom-wide ribbon.
The absence of bandgap is evident and near the Fermi level, where the bands meet, a large density of states is observed.
The probability density of the degenerate states closing the gap near the edge of the Brillouin zone ($k = \pi/a_0$) is plotted in Fig.~\ref{f:zigzag_08_valence}.
Because their wavefunctions differ only in phase, the degenerate states have equal probability density.
As expected from zigzag-edge ribbons, we observe strong localization near the edges of the ribbons.
 
\begin{figure}[h!]
  \includegraphics{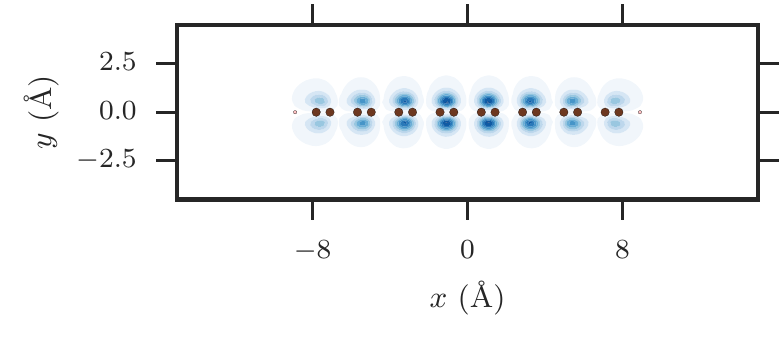}%\\
  \caption{Squared amplitude of the first valence state of a 8 C wide zigzag ribbon at $k=\pi/a_0$.}
  \label{f:zigzag_08_valence_lowk}
\end{figure}

As exemplified in Fig.~\ref{f:zigzag_08_valence_lowk}, valence and conduction states closer to the center of the first Brillouin zone, i.e. further from the Fermi level, regain their bulk-like character.
%as indicated in figures~\ref{f:zigzag_08_valence_lowk} and \ref{f:zigzag_08_conduction_lowk}.
%\begin{figure}[h!]
%  \includegraphics{assets/zigzag_08_conduction_z_lowk.pdf}%\\
%  \caption{Squared amplitude of the first conduction state of a 8 C wide zigzag ribbon at $k=\pi/2a_0$.}
%  \label{f:zigzag_08_conduction_lowk}
%\end{figure}
However, because the high density of states at the Fermi level has the ability to pin the Fermi level, the edge states will be responsible for most of the transport characteristics within a wide range of parameters.

\section{Tunneling current}
\label{tunneling_current}

Using the Bardeen transfer Hamiltonian method, we determine the direct tunneling current from the electronic structure calculations.
In order to keep the parameter space reasonable we have considered perpendicular ribbons with a spacing of $1$ nm vacuum between them. 
While vacuum is used in this model, a dielectric would offer better coupling and higher tunnel currents for the same inter-ribbon distance.
We show results for the tunneling current between two armchair-edge ribbons and also between an armchair-edge and a zigzag-edge ribbon.
In each case we calculated the current for increasing ribbon sizes and different values of the applied bias voltage.

\subsection{Armchair-to-armchair}

\begin{figure}[h!]
  \includegraphics[width=0.4\textwidth]{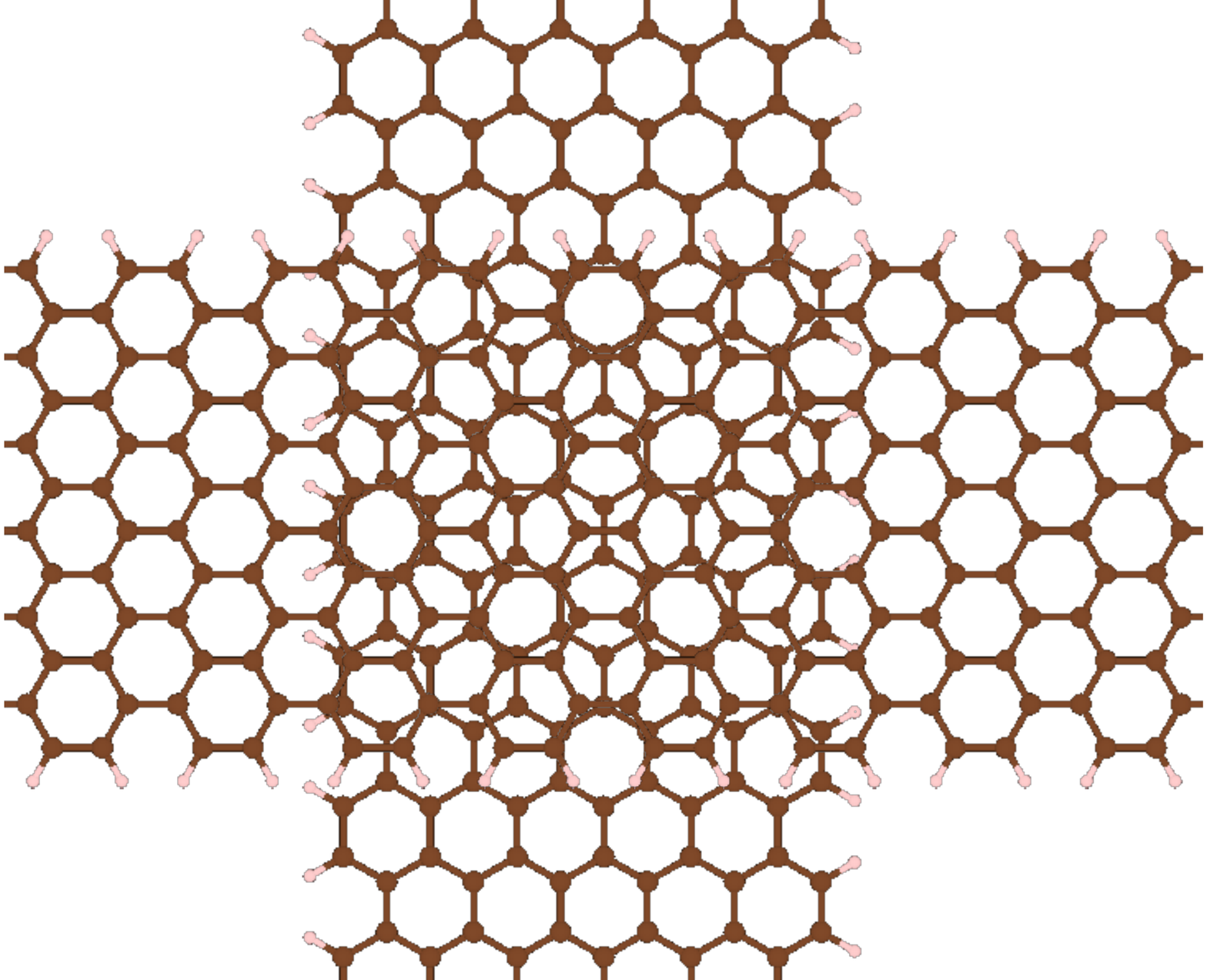}
  \caption{Top view of an armchair-to-armchair tunneling structure. The armchair ribbons both have a width of $12$ carbon lines. The edges are hydrogen terminated.}
  \label{f:aa-top-view}
\end{figure}

We consider tunneling between two armchair-edge ribbons first.
The $90^\circ$ rotation between the ribbons results in incommensurate lattices as can be seen in Fig.~\ref{f:aa-top-view} where the top view of two ribbons placed over each other shows part of a Moir\'e pattern.
Because this configuration does not have a natural alignment, we have to choose how to align the ribbons as, in general, this will influence the tunneling current.
We have aligned the center of the center-most sextet of both ribbons, allowing to compare between increasing ribbon widths while maintaining the same alignment.

\begin{figure}[h!]
  \includegraphics{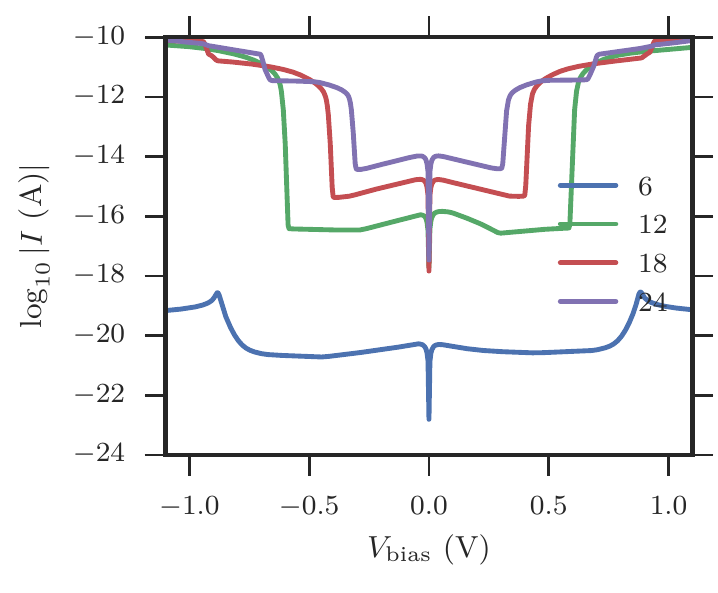}
  \caption{Tunneling current between intrinsically doped armchair ribbons with applied bias $V_{\rm ds}$ for increasing width. The legend shows the width in number of carbon lines, all ribbons here are part of the $3p$ category with intermediate bandgap, as explained in section~\ref{electronic_structure}.}
  \label{f:aa-currents}
\end{figure}

In Fig.~\ref{f:aa-currents} we plot the tunneling current versus applied bias voltage for increasing ribbon width without any doping.
We observe a similar behaviour for all ribbon widths; a negligible current flows at low applied bias followed by a sharp increase in current as the applied bias is raised.
Close inspection reveals that the onset of the current increase occurs at a bias voltage equal to the bandgap.

We explain this behavior by considering the density-of-states of both ribbons that was shown in Fig.~\ref{f:armchair_16_bsdos}, and its occupation at different bias for the configuration.
For intrinsic armchair-edge ribbons the Fermi level lies approximately in the middle of the bandgap, the valence band is therefore almost completely filled, while the conduction band is empty.
Tunneling current can therefore only take place between the valence and conduction band of different ribbons.
At low bias, the bandgap of both ribbons blocks the transfer of carriers from one to the other. 
Due to the thermal distribution of the electrons, a very small tunneling current is always present when bias is applied.
However, because the bandgaps block a large part of the tail of the distribution, a sub-threshold like region of exponentially increasing current is not present.
This is similar to the effect of energy filtering in tunneling field-effect transistors\cite{Ionescu:2011dr}.

When the bias voltage is increased to the value of the bandgap, the first conduction band of one ribbon starts overlapping with the valence band of the other ribbon.
Both the valence and conduction band have a high DoS due to the 1D nature of the graphene ribbons. 
Because the valence band is filled, the corresponding electrons tunnel to the empty states in the conduction band of the other ribbon.
This results in a sharp onset of current.
Further increasing the bias beyond this point causes the overlap of the bands in a larger energy range, but this increase is counteracted by the magnitude of the overlap as the 1D peaks of the DoS become misaligned, resulting in a quasi-constant current for increasing bias, until another band is crossed and a new step emerges.

The working principle of these armchair-to-armchair inter-ribbon tunneling structures depends highly on the DoS and the occupation of the ribbon states, and not on the transmission coefficients which only serve as a current limiting factor.
Physically, this corresponds to a strong dependence on the ribbon width and on its doping, but not on the relative orientation of the ribbons. 
Furthermore, while the complete structure is not symmetric with respect to interchanging the terminals, we do observe near-perfect symmetry in the current when inverting the bias. This further indicates the relatively small importance of the alignment at these small dimensions.

\subsection{Zigzag-to-armchair}

\begin{figure}[h!]
  \includegraphics[width=0.4\textwidth]{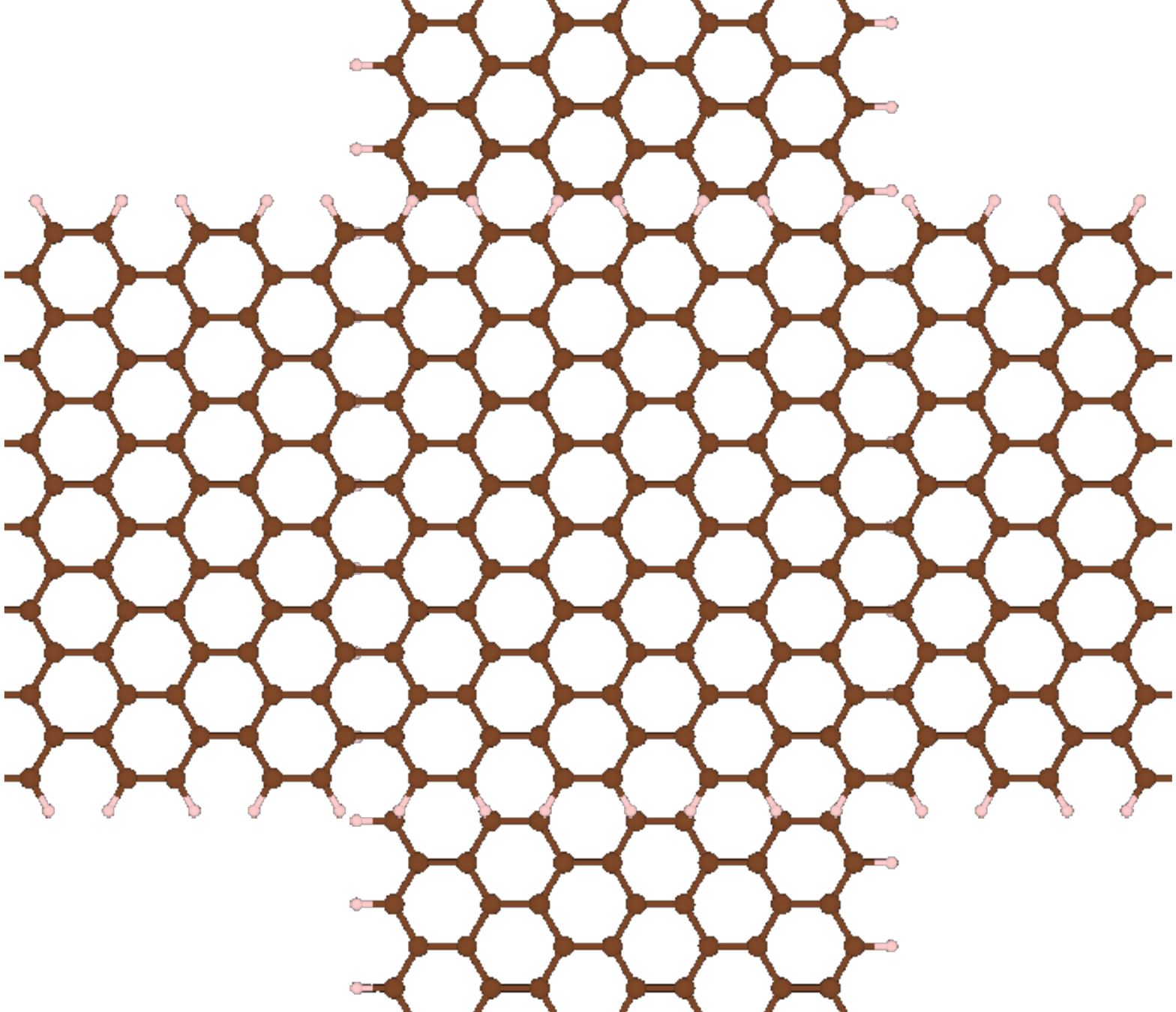}
  \caption{Top view of a zigzag-to-armchair tunneling structure. The zigzag ribbon is $7$ carbon chains wide, whereas the armchair ribbon has a width of $14$ carbon lines. Carbon is depicted with gray, while hydrogen termination is off-white.}
  \label{f:za-top-view}
\end{figure}

In the case of tunneling between an armchair-edge ribbon and a zigzag-edge ribbon, the $90^\circ$ rotation ensures that the lattices are oriented in the same way.
We align the ribbons in the natural way where the atoms are right on top of each other as shown in Fig.~\ref{f:za-top-view}.

\subsubsection{Tunneling between intrinsic ribbons}

\begin{figure}[h!]
  \includegraphics{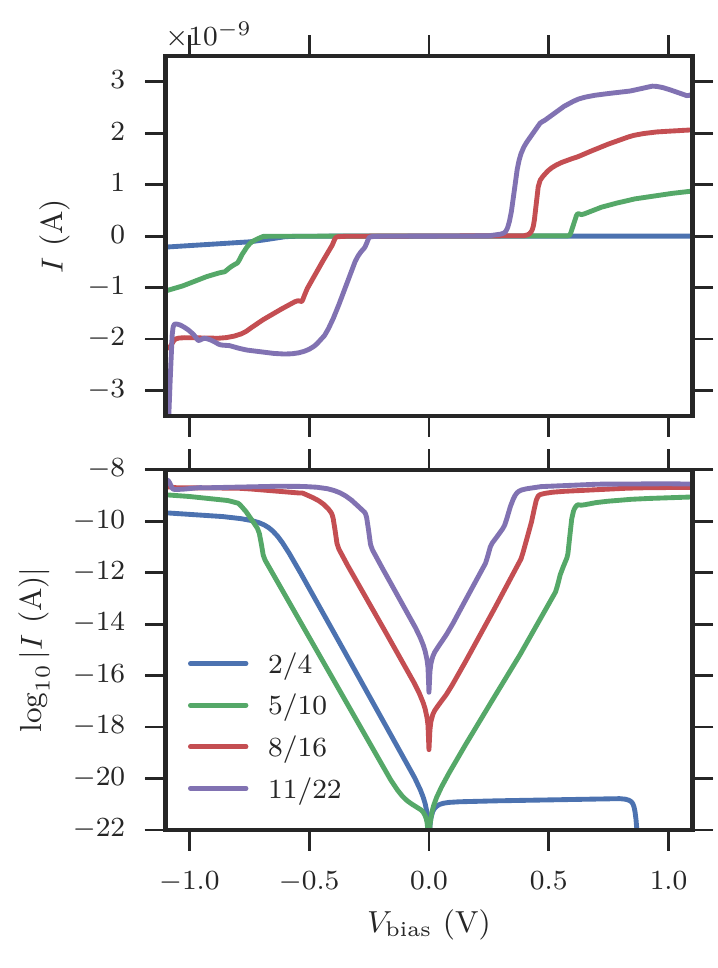}
  \caption{Tunneling current between intrinsically doped armchair and zigzag ribbons with applied bias $V_{\rm ds}$ for increasing ribbon widths.}
  \label{f:za-currents}
\end{figure}

Starting with undoped ribbons, we show the current versus applied bias for different ribbon widths in Fig.~\ref{f:za-currents}.
In contrast to the armchair-to-armchair tunneling, we observe a ``sub-threshold'' region of exponential increase in current with increasing bias.
At higher bias, the current sharply increases to a plateau as observed in the armchair-to-armchair tunneling case.
The current increases approximately at a bias halfway the bandgap.

The ``sub-threshold'' region is a direct result of the absence of a bandgap in the zigzag-edge ribbon.
As the bias is increased, carriers from the thermal tail of the distribution in the zigzag ribbon tunnel to the valence or conduction band (depending on the polarity of the bias) of the armchair ribbon.
This causes the typical exponential increase of the current with increasing bias.
At a bias of about half the bandgap the peaked DoS at the Fermi level of the zigzag ribbon overlaps with the conduction or valence band of the armchair ribbon, this causes an abrupt increase in current just as in the armchair-to-armchair tunneling case.

In section~\ref{electronic_structure} we highlighted the role played by the edge states of the zigzag ribbon near the Fermi level.
Because the tunneling current is determined by these states we expect a linear scaling when increasing the size of both ribbons instead of quadratic scaling as for the ribbon overlap.
Indeed, in Fig.~\ref{f:za-currents} we observe this linear scaling, which is indicative of the edge states carrying the direct tunneling current.

The zigzag-armchair structure shows strong ambipolar behaviour with quasi-symmetric tunneling current with respect to the reversal of the applied bias.
This is unexpected in a non-symmetric structure.
The ambipolarity is a direct result of the high DoS from the edge states near the Fermi level in the zigzag ribbon and the symmetric bandstructure of the armchair ribbon around the bandgap.
We note a deviation from the ambipolarity at the smallest dimensions, namely a two carbon chains wide zigzag and a 4 carbon lines wide armchair ribbon.
In this case, a positive bias does not increase tunneling current at all because the parity of the overlapping states forbids electron transfer, i.e. the matrix element is zero.

\subsubsection{Tunneling between doped ribbons}

As we have demonstrated in the preceding sections, the tunneling current is mainly determined by the DoS and the occupation, which are both heavily influenced by doping.
To account for doping effects, we assume a fixed amount of doping charge, resulting in a fixed concentration of free carriers in the conduction or valence band.
We do not specifically consider the possible physical origin of the doping such as chemical doping, or dynamically by electrostatic gating.
The latter is interesting from an application point of view as it allows gate control over the effects described here.
In our model, the doping simply determines the Fermi level and the occupation of the 1D DoS peaks.
The Fermi level in the zigzag-edge ribbons is insensitive to changes to the doping level because of the pinning effect of the large DoS.
In contrast, armchair-edge ribbons are sensitive to even low levels of doping due to a large bandgap and 1D DoS at the band edge.

\begin{figure}[h!]
  \includegraphics{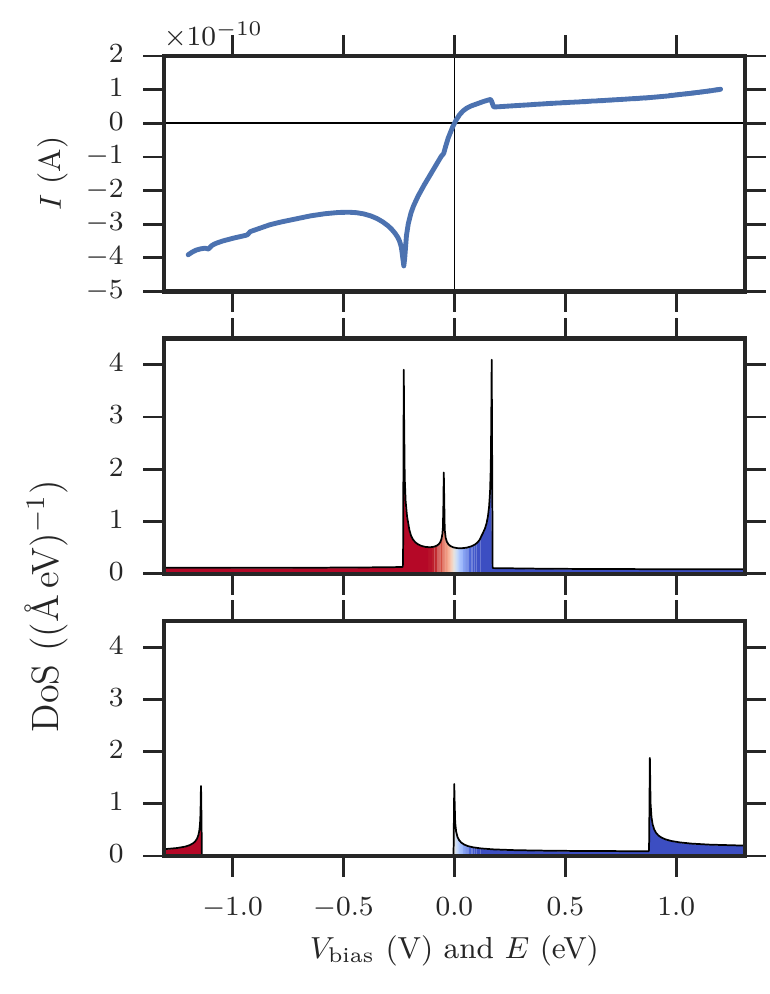}
  \caption{Tunneling current between n-doped armchair and zigzag ribbons with applied bias $V_{\rm ds}$ (top). The corresponding density of states (DoS) of the zigzag ribbon (middle) and armchair ribbon (bottom) with Fermi level at $0$ eV. The DoS is shaded with the occupation probability, where red is filled and blue is empty.}
  \label{f:za-currents-n}
\end{figure}

We show the results for tunneling between $n$-doped zigzag and armchair-edge ribbons in Fig.~\ref{f:za-currents-n}.
From the DoS of both ribbons we clearly see the effect of doping.
As expected, we observe only a negligible shift in the Fermi level of the zigzag-edge ribbon compared to the intrinsic case, while the same doping level has shifted the Fermi level of the armchair ribbon completely to the conduction band.

With $n$-type doping at negative bias the current increases towards a resonant peak, followed by a region of negative differential resistance (NDR) where the current decreases.
The resonant tunneling peak coincides with the energetic difference between the 1D DoS peaks of the zigzag and armchair-edge ribbons.
At this bias the huge 1D DoS of both ribbons are aligned causing a large number of states to tunnel, yielding a high current.
If the peaks are not aligned, the number of overlapping states is lower, reducing the current, even at an absolute bias higher than that where the resonance occurs.
Even a slight deviation from alignment of the DoS peaks causes a sudden decrease in available tunneling possibilities, resulting in a sharp resonance peak.
In this structure, the resonance is effectively caused by the DoS of both reservoirs, as opposed to the resonance in a `standard' resonant tunneling diode where quasi-bound states between two barriers provides the peaked DoS required for the resonant current flow.
The doping levels determine the bias required for resonance by moving the DoS peaks, but it also influences the peak-to-valley ratio.
Higher doping levels cause depletion or accumulation of electrons in the 1D DoS peaks which reduces the overall tunneling current.

Within the Bardeen transfer Hamiltonian method the tunneling current between two coupled 1D DoS can diverge due to the $1/E$ divergence appearing in the energy integrals.
However, the current will be limited by the ballistic limit of the ribbons themselves that have to supply the electrons to the tunneling area.
The proper treatment of this interplay is not possible using perturbative methods and requires a description of the full system non-perturbatively.
Methods such as non-equilibrium Green's functions (NEGF) and quantum transmitting boundary conditions (QTBM) would provide a better description at a higher cost of computational resources.
Furthermore, using these advanced methods incurs a loss of the straightforward interpretation as obtained from the perturbative Bardeen approach.

At positive bias we also observe an abrupt decrease in current at a bias equal to the energy difference between the 1D DoS peak of the zigzag ribbon above the Fermi level and the peak of the armchair ribbon conduction band.
The degenerate doping of the armchair ribbon is responsible for the current between conduction band states of both ribbons.
Initially, at low bias, the current is significant due to the high DoS near the Fermi level of the zigzag ribbon.
The abrupt decrease of current occurs when the high DoS of the zigzag ribbon no longer overlaps with the DoS peak of the armchair ribbon.
Essentially, this is the inverse of the effect we have observed in armchair to armchair tunneling.

\section{Conclusions}
\label{conclusions}

We used empirical pseudopotentials to implement a highly scalable, atomistic electronic structure solver based on the RMM-DIIS eigenvalue algorithm.
With the Bardeen transfer Hamiltonian we obtained direct tunneling current between crossed graphene ribbons from the obtained atomistic electronic structures of the individual ribbons.
The Bardeen transfer Hamiltonian method, relying on an atomistic band structure method, such as pseudopotential calculations, is an excellent way to accurately describe interlayer tunneling.

We have studied the tunneling current for different configurations of armchair and zigzag-edge ribbons and have found the density of states and occupation of states to be the most significant parameters to determine the current.
The one-dimensional character in the density of states of the ribbons gives rise to either resonant tunneling or a step-like increase in current.
A step like increase is caused by the overlap of two band-edges from different ribbons with band curvatures of opposite sign.
We have shown this effect in the armchair-to-armchair tunneling structure, where the first conduction band of one ribbon overlaps the first valence band of the second ribbon.
On the other hand, resonant tunneling is caused by the alignment of two band-edges with band curvature of the same sign as we observed in tunneling between zigzag and armchair-edge ribbons.
Furthermore, through control of the doping level, for example by electrostatic gating, we can control both the occupation of the states and the position of the peaks in the DoS relative to the Fermi level.
Thanks to the possibility of gate control combined with the very abrupt current steps, these one-dimensional structures may prove promising candidates for a next generation of steep sub-threshold slope devices.

Because the effects of resonance and stepwise increases in current are features of the dimensionality of the ribbons, as reflected in the one-dimensional DoS, we can generalize our conclusions.
Any two conductors that exhibit a one-dimensional DoS and that are brought within tunneling range can exhibit these effects, provided that there are no forbidden transitions due to symmetry.
Conductors that exhibit a strong one-dimensional signature in their DoS include ribbons made from alternative two dimensional materials such as transition metal dichalcogenides (TMDs), but also carbon nanotubes or even very thin nanowires.
Because most of these structures have a bandgap, they will resemble tunneling between armchair-edge graphene nanoribbons at small bias, resulting in sharp turn-on of tunneling current due to the peaked density-of-states near the band gap.
Lastly, quantitative results for these structures can be obtained using the same numerical technique based on emperical pseudopotentials and the Bardeen transfer Hamiltonian we have used in this work to model the inter-ribbon tunneling in graphene.

%\section*{Acknowledgments}

\bibliography{main}

\end{document}